\documentclass{elsart3}
\usepackage{graphics}
\usepackage{epsfig}
\usepackage{amssymb}
\usepackage{cite}

\begin{document}
  \newcommand{\greeksym}[1]{{\usefont{U}{psy}{m}{n}#1}}
  \newcommand{\umu}{\mbox{\greeksym{m}}}
  \newcommand{\udelta}{\mbox{\greeksym{d}}}
  \newcommand{\uDelta}{\mbox{\greeksym{D}}}
  \newcommand{\uOmega}{\mbox{\greeksym{W}}}
  \newcommand{\uPi}{\mbox{\greeksym{P}}}
  \newcommand{\ualpha}{\mbox{\greeksym{a}}}
  \begin{frontmatter}


\title{Observation, modeling, and temperature dependence of doubly peaked electric fields in irradiated silicon pixel sensors}

\author[jhu]{M.~Swartz},
\author[uniz]{V.~Chiochia},
\author[uniz]{Y.~Allkofer},
\author[purdue]{D.~Bortoletto},
\author[miss]{L.~Cremaldi},
\author[basel]{S.~Cucciarelli},
\author[uniz,psi]{A.~Dorokhov},
\author[uniz,psi]{C.~H\"ormann},
\author[jhu]{D.~Kim},
\author[basel]{M.~Konecki},
\author[psi]{D.~Kotlinski},
\author[uniz,psi]{K.~Prokofiev},
\author[uniz]{C.~Regenfus},
\author[psi]{T.~Rohe},
\author[miss]{D.~A.~Sanders},
\author[purdue]{S.~Son},
\author[uniz]{T.~Speer}

\address[jhu]{Johns Hopkins University, Baltimore, MD 21218, USA}
\address[uniz]{Physik Institut der Universit\"at Z\"urich-Irchel, 8057 Z\"urich, Switzerland}
\address[purdue]{Purdue University, West Lafayette, IN 47907, USA}
\address[miss]{University of Mississippi, University, MS 38677, USA}
\address[basel]{Institut f\"ur Physik der Universit\"at Basel, 4056 Basel, Switzerland}
\address[psi]{Paul Scherrer Institut, 5232 Villigen PSI, Switzerland}
\begin{abstract}
We show that doubly peaked electric fields are necessary to describe grazing-angle charge collection measurements of irradiated silicon pixel sensors.  A model of irradiated silicon based upon two defect levels with opposite charge states and the trapping of charge carriers can
be tuned to produce a good description of the measured charge collection profiles
in the fluence range from $0.5\times$10$^{14}$~n$_{\rm eq}$/cm$^2$
to $5.9\times$10$^{14}$~n$_{\rm eq}$/cm$^2$.  The model correctly predicts the variation in the profiles as the temperature is changed from $-10^\circ$C to $-25^\circ$C.  The measured charge collection profiles are inconsistent with the linearly-varying electric fields predicted by the usual description based upon a uniform effective doping density.  This observation calls into question the practice of using effective doping densities to characterize irradiated silicon.
\end{abstract}

\end{frontmatter}

\section{Introduction}
A silicon pixel detector~\cite{CMS_Tracker_TDR} is currently being developed for the CMS experiment at the CERN Large Hadron Collider (LHC).  The detector will be a key component in the reconstruction of primary and secondary vertices in the particularly harsh LHC environment characterized by large track multiplicities and high radiation backgrounds.
The innermost layer, located at only 4 cm from the beam line, is expected to be exposed to a 1 MeV neutron equivalent fluence of $3 \times 10^{14}$~n$_{\rm eq}$/cm$^2$ per year at full luminosity.

The response of the silicon sensors during the detector
operation is of great concern. It is well understood that the intra-diode electric fields in these detectors vary linearly in depth reaching a maximum value at the p-n junction.  The linear behavior is a consequence of a uniform space charge density, $N_{\rm eff}$, caused by thermally ionized impurities in the bulk material.  It is well known that the detector characteristics are affected by radiation exposure, but it is generally assumed that the same picture is valid after irradiation.  In fact, it is common to characterize the effects of irradiation in terms of a varying effective uniform charge density.
In~\cite{Chiochia:2004qh} we have proved that this picture does not provide a good description of
irradiated silicon pixel sensors. In addition, it was shown that it is possible to adequately describe
the charge collection characteristics of a heavily irradiated silicon detector in terms 
of a tuned double junction model which produces a doubly peaked electric field profile across the sensor.  The modeling is supported by the evidence of doubly peaked electric fields obtained directly from beam
test measurements and presented in~\cite{Dorokhov:2004xk}.  The dependence of the modeled trap concentrations upon fluence was presented in \cite{Chiochia:2005ag}.
In this paper, we summarize the previous results and investigate the temperature dependence of the model.

This paper is organized as follows: Section~\ref{sec:technique} describes
the experimental setup, Section~\ref{sec:simulation} describes the carrier transport 
simulation used to interpret the data. The tuning of the double junction model and its resulting predictions are discussed in Section~\ref{sec:data_analysis}. The temperature dependence of the data and model are summarized in Section~\ref{sec:temperature}.  The conclusions are given in Section~\ref{sec:conclusions}.
\section{Experimental setup\label{sec:technique}}
The measurements were performed in the H2 beam  line of the CERN SPS in 2003/04 using 150-225 GeV pions.  The beam test apparatus is described in~\cite{Dorokhov:2003if}.
A silicon beam telescope \cite{Amsler:2002ta} consisted of four modules each containing two 300~$\umu$m thick single-sided silicon detectors with a strip pitch of 25 $\umu$m and readout pitch of 50~$\umu$m.  The two detectors in each module were oriented to measure horizontal and vertical impact coordinates.  A pixel hybrid detector was mounted between the second and third telescope modules on a cooled rotating stage.  A trigger signal was generated by a silicon
PIN diode. The analog signals from all detectors were digitized in a VME-based readout system by two CAEN (V550) ADCs and one custom-built flash ADC. 
The entire assembly was located in an open-geometry 3T Helmholtz magnet that produced a magnetic field either parallel or orthogonal to the beam. 
The temperature of the tested sensors was controlled with a Peltier cooler that was capable of operating down to -30$^\circ$C.  The telescope information was used to reconstruct the trajectories of individual beam particles and to achieve a precise determination of the particle hit position
in the pixel detector.  The resulting intrinsic resolution of the beam telescope was about 1~$\umu$m.

The prototype pixel sensors are so-called ``n-in-n'' devices: they are designed to collect charge from n$^+$ structures implanted into n--bulk silicon using p-spray isolation. All test devices were 22$\times$32 arrays of 125$\times$125~$\mu$m$^2$ pixels that were fabricated by CiS.  The substrate, produced by Wacker, was 285~$\mu$m thick, n-doped, diffusively-oxygenated float zone silicon of orientation $\langle111\rangle$, resistivity 3.7~k$\Omega\cdot$cm and oxygen concentration in the order of $10^{17}$ cm$^{-3}$.  Individual sensors were diced from fully processed wafers after the deposition of under-bump metalization and indium bumps.  A number of sensors were irradiated at the CERN PS with 24 GeV protons. The irradiation was performed without cooling or bias. The delivered proton fluences scaled to 1 MeV neutrons by the hardness factor 0.62~\cite{Lindstrom:2001ww} were $0.5\times$10$^{14}$~n$_{\rm eq}$/cm$^2$, $2\times$10$^{14}$~n$_{\rm eq}$/cm$^2$ and $5.9\times$10$^{14}$~n$_{\rm eq}$/cm$^2$. All samples were annealed for three days at 30$^\circ$C. In order to avoid reverse annealing, the sensors were stored at -20$^\circ$C after irradiation and kept at room temperature only for transport and bump bonding. All sensors were bump bonded to PSI30/AC30 readout chips \cite{Meer:PSI30} which allow analog readout of all 704 pixel cells without zero suppression.  The PSI30 settings were adjusted to provide a linear response to input signals ranging from zero to more than 30,000 electrons.
\section{Sensor simulation\label{sec:simulation}}
The interpretation of the test beam data relies upon a detailed sensor simulation that includes the modeling of irradiation effects in silicon.
The simulation, {\sc pixelav}~\cite{Swartz:2003ch,Swartz:CMSNote,Chiochia:2004qh}, incorporates the following elements: an accurate model of charge deposition by primary hadronic tracks (in particular to model delta rays); a realistic 3-D intra-pixel electric field map; an established model of charge drift physics including mobilities, Hall Effect, and 3-D diffusion; a simulation of charge trapping and the signal induced from trapped charge; and a simulation of electronic noise, response, and threshold effects.  The intra-pixel electric field map was generated using {\sc tcad} 9.0 \cite{synopsys} to simultaneously solve Poisson's Equation, the carrier continuity equations, and various charge transport models.  A final simulation step reformatted the simulated data into test beam format so that it could be processed by the test beam analysis software.

The effect of irradiation was implemented in the {\sc tcad} simulation by including
two defect levels in the forbidden silicon bandgap with opposite
charge states and trapping of charge carriers. The model, similar to one proposed in~\cite{Eremin:2002wq},
is based on the Shockley-Read-Hall (SRH) statistics 
and produces an effective space charge density $\rho_\mathrm{eff}$ from the trapping 
of free carriers in the leakage current.  
The effective charge density is related to the occupancies and densities of traps as follows,
\begin{equation}
\rho_\mathrm{eff} = e\left[N_Df_D-N_Af_A\right] + \rho_\mathrm{dopants} 
\end{equation}
where: $N_D$ and $N_A$ are the densities of donor and acceptor trapping states, respectively; $f_D$ and $f_A$ are the occupied fractions of the donor and acceptor states, respectively, and $\rho_\mathrm{dopants}$ is the charge density due to ionized dopants (describes the resistivity of the material before irradiation).  The donor and acceptor occupancies are related to the trap parameters by standard SRH expressions
\begin{eqnarray}
f_D &=& \frac{v_h\sigma^D_hp+v_e\sigma^D_en_ie^{E_D/kT}}{v_e\sigma^D_e(n+n_ie^{E_D/kT})+v_h\sigma^D_h(p+n_ie^{-E_D/kT})}   \nonumber \\
\\
f_A &=& \frac{v_e\sigma^A_en+v_h\sigma^A_hn_ie^{-E_A/kT}}{v_e\sigma^A_e(n+n_ie^{E_A/kT})+v_h\sigma^A_h(p+n_ie^{-E_A/kT})}  \nonumber
\end{eqnarray}
where: $v_e$ and $v_h$ are the thermal speeds of electrons and holes, respectively; $\sigma_e^D$, $\sigma_h^D$ are the electron and hole capture cross sections for the donor trap; $\sigma_e^A$, $\sigma_h^A$ are the electron and hole capture cross sections for the acceptor trap; $n$, $p$ are the densities of free electrons and holes, respectively; $n_i$ is the intrinsic density of carriers; $E_D$, $E_A$ are the activation energies (relative to the mid-gap energy) of the donor and acceptor states, respectively.  Note that the single donor and acceptor states model the effects of many physical donor and acceptor states making the two-trap model an ``effective theory''.  

\begin{figure}[hbt]
%
%
  \begin{center}
    \resizebox{0.8\linewidth}{!}{\includegraphics{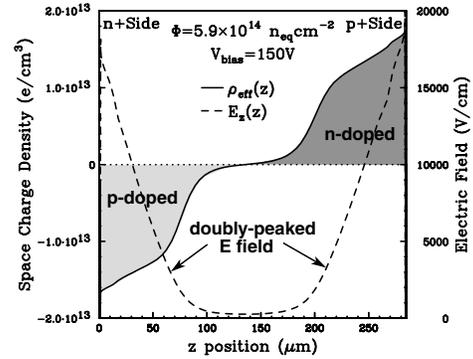}}
  \caption{The space charge density (solid line) and electric field (dashed line) at $T=-10^\circ$C as functions of depth in a two-trap double junction model tuned to reproduce the $\Phi=5.9\times10^{14}$n$_{\rm eq}$cm$^{-2}$ charge collection data at 150V bias.}
  \label{fig:evl_sketch} 
  \end{center}
\end{figure}

The physics of the model is illustrated in Fig.~\ref{fig:evl_sketch}. The space charge density and electric field are plotted as functions of depth $z$ for a model tuned to reproduce the $\Phi=5.9\times10^{14}$n$_{\rm eq}$cm$^{-2}$ charge collection data at 150V bias.  The SRH process produces electron-hole pairs more or less uniformly across the thickness of the sensor.  As the electrons drift to the n+ implant, the total electron current increases as $z$ decreases.  The hole current similarly increases with increasing $z$.  Trapping of the mobile carriers produces a net positive space charge density near the p$^+$ backplane and a net negative space charge density near the n$^+$ implant.  Since positive space charge density corresponds to n-type doping and negative space charge corresponds to p-type doping, there are p-n junctions at both sides of the detector.  The electric field in the sensor follows from a simultaneous solution of Poisson's equation and the continuity equations.  The resulting $z$-component of the electric field varies with an approximately quadratic dependence upon $z$ having a minimum at the zero of the space charge density and maxima at both implants.  A more detailed description of the double junction model and its implementation
can be found in~\cite{Chiochia:2004qh}.
\section{Model tuning and results\label{sec:data_analysis}}
\begin{figure}[hbt]
%
%
  \begin{center}
    \resizebox{\linewidth}{!}{\includegraphics{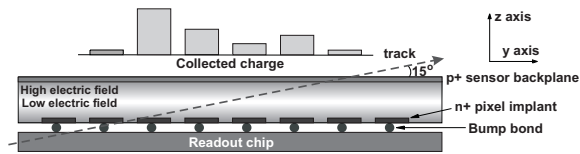}}
  \caption{The grazing angle technique for determining charge collection profiles.  The charge measured by each pixel along the $y$ direction samples a different depth $z$ in the sensor.}
  \label{fig:fifteen_deg} 
  \end{center}
\end{figure}

%
Charge collection across the sensor bulk was measured using the ``grazing angle technique''~\cite{Henrich:CMSNote}.  As is shown in Fig.~\ref{fig:fifteen_deg}, the surface of the test sensor is oriented by a small angle (15$^\circ$) with respect to the pion beam.  Several samples of data were collected with zero magnetic field and at temperature of $-10^\circ$C and $-25^\circ$C.  The charge measured by each pixel along the $y$ direction samples a different depth $z$ in the sensor.  Precise entry point information from the beam telescope is used to produce finely binned charge collection profiles.

\begin{figure}[hbt]
  \begin{center}
    \resizebox{\linewidth}{!}{\includegraphics{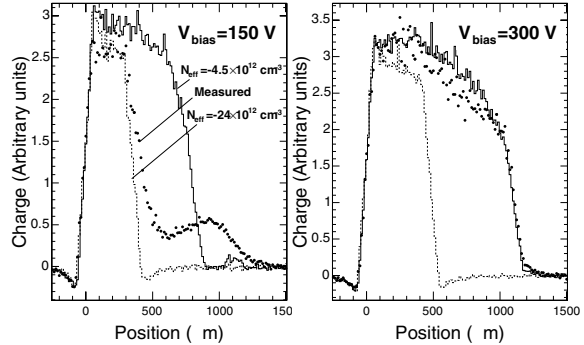}}
  \caption{The measured and simulated charge collection profiles for a sensor at $T=-10^\circ$C irradiated to a fluence of $\Phi=5.9\times10^{14}$~n$_{\rm eq}/$cm$^{2}$.  The profiles measured at bias voltages of 150V and 300V are shown as solid dots.  The full histograms are the simulated profiles for a constant effective doping  $N_{\rm eff}=4.5\times10^{12}$~cm$^{-3}$ of acceptor impurities.  The dashed histograms are the simulated profiles for a constant effective doping  $N_{\rm eff}=24\times10^{12}$~cm$^{-3}$.}
  \label{fig:strawmen} 
  \end{center}
\end{figure}

The charge collection profiles for a sensor irradiated to a fluence of $\Phi=5.9\times10^{14}$~n$_{\rm eq}/{\rm cm}^{2}$ and operated at a temperature of $-10^\circ$C and bias voltages of 150V and 300V are presented in Fig~\ref{fig:strawmen}.  The measured profiles are shown as solid dots and the simulated profiles are shown as histograms.  In order to investigate the applicability of the traditional picture of type-inverted silicon after irradiation, the simulated profiles were generated with electric field maps corresponding to two different effective densities of acceptor impurities.  The full histograms are the simulated profile for $N_{\rm eff}=4.5\times10^{12}$~cm$^{-3}$.  Note that the 300V simulation reasonably agrees with the measured profile but the 150V simulation is far too broad.  The dashed histograms show the result of increasing $N_{\rm eff}$ to $24\times10^{12}$~cm$^{-3}$.  At this effective doping density, the width of the simulated peak in the 150V distribution is close to correct but it does not reproduce the ``tail'' observed in the data at large $y$.  The 300V simulated distribution is far too narrow and the predicted charge is lower than the data (note that the profiles are absolutely normalized).  It is clear that a simulation based upon the standard picture of a constant density of ionized acceptor impurities cannot reproduce the measured profiles.

\begin{figure}[hbt]
  \begin{center}
    \resizebox{0.9\linewidth}{!}{\includegraphics{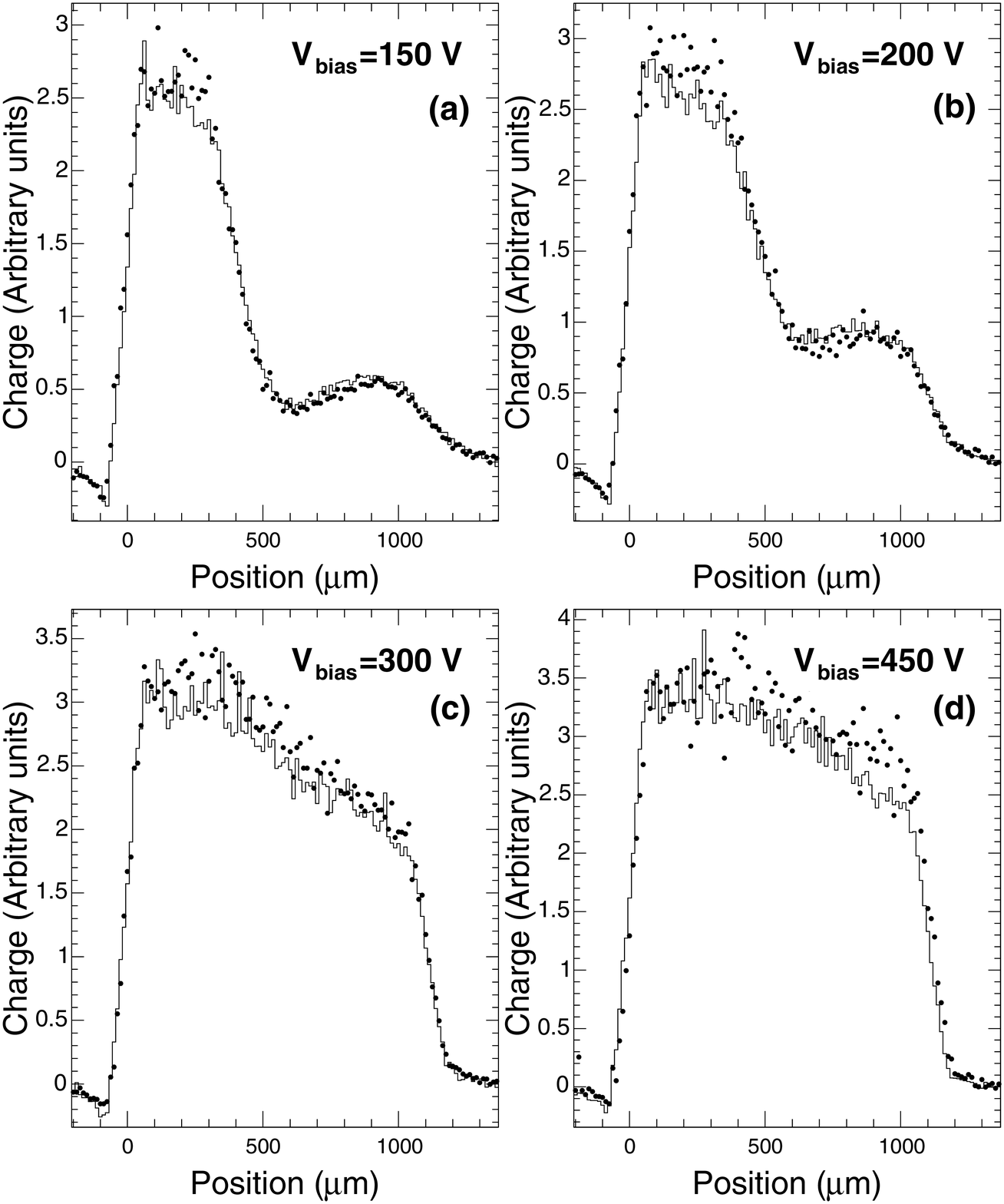}}
  \caption{The measured charge collection profiles at a temperature of $-10^\circ$C and bias voltages of 150V, 200V, 300V, and 450V are shown as solid dots for a fluence of $5.9\times10^{14}$~n$_{\rm eq}/$cm$^{2}$.  The two-trap double junction simulation is shown as the solid histogram in each plot.}
  \label{fig:dj44} 
  \end{center}
\end{figure}

\begin{figure}[hbt]
  \begin{center}
    \resizebox{0.55\linewidth}{!}{    
      \includegraphics[angle=90]{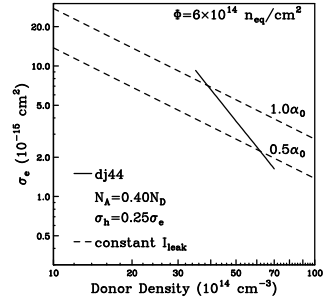} 
	      }
  \end{center}
  \caption{The allowed region in the $N_D$-$\sigma_e$ space for the best fit $5.9\times10^{14}$~n$_{\rm eq}/$cm$^{2}$ model is shown as the solid line.  Contours of constant leakage current are shown as dashed curves and are labeled in terms of the corresponding damage parameter $\alpha$ where $\alpha_0=4\times10^{-17}\ \mathrm{A/cm}$ is the expected leakage current \cite{mfl}.    \label{fig:dj35_parameters} }
\end{figure}

The same measured profiles and those from bias voltages of 200V and 450V are shown in Fig.~\ref{fig:dj44}.  They are compared with simulations based upon the electric field produced by the two trap model.
The model has six free parameters 
($N_D$, $N_A$, $\sigma_e^D$, $\sigma_h^D$, $\sigma_e^A$, $\sigma_h^A$) 
that can be adjusted.  The activation energies are kept fixed to the values of~\cite{Eremin:2002wq}: $E_D=E_V+0.48$~eV, $E_A=E_C-0.525$~eV where $E_V$ and $E_C$ are the energies of the valence and conduction band edges.  The electric field map produced by each {\sc tcad} run is input into  {\sc pixelav}.  The electron and hole trapping rates, $\Gamma_e$ and $\Gamma_h$, are also inputs to  {\sc pixelav} and are treated as constrained parameters.  Although they have been measured \cite{kramberger}, they are uncertain at the 20\% level due to the fluence uncertainty and possible annealing of the sensors.  They are therefore allowed to vary by as much as $\pm$20\% from their nominal values. The donor concentration of the starting material is set to $1.2\times10^{12}$~cm$^{-3}$ corresponding to a full depletion voltage of about 70~V for an unirradiated device.  Because each model iteration took approximately two days, it was not possible to use standard statistical fitting techniques.
The parameters of the double junction model were systematically varied and the agreement 
between measured and simulated charge collection profiles was judged subjectively.  The ``best fits'' shown in this paper are probably not true likelihood minima and the calculation of eight parameter error matrices is beyond available computational resources.
Adequate agreement was achieved by setting the ratio of the common hole and electron cross sections $\sigma_h/\sigma_e$ to 0.25 and the ratio of the acceptor and donor densities $N_A/N_D$ to 0.40.  There is a range of parameters in the $N_D$-$\sigma_e$ space that produces reasonable agreement with the measured profiles.  The range is shown in Fig.~\ref{fig:dj35_parameters}a as the solid line in the logarithmic space.  If the donor density becomes too small ($N_D<20\times10^{14}$~cm$^{-3}$), the 150V simulation produces too much signal at large $z$.  If the donor density becomes too large ($N_D>50\times10^{17}$~cm$^{-3}$), the 300V simulation produces insufficient signal at large $z$.  Since the simulated leakage current varies as $I_\mathrm{leak}\propto\sigma_e N_D$, different points on the allowed solid contour correspond to different leakage current.  Contours of constant leakage current are shown as dashed curves and are labeled in terms of the corresponding damage parameter $\alpha$ where $\alpha_0=4\times10^{-17}\ \mathrm{A/cm}$ is the expected leakage current \cite{mfl}.  It is clear that the simulation can accommodate the expected leakage current which is smaller than the measured current by a factor of three.  The same choice of parameters can also account for the observed rate of signal trapping \cite{Chiochia:2004qh}.

The simulation describes the measured charge collection profiles well both
in shape and normalization. 
The ``wiggle'' observed at low bias voltages is a signature of the doubly peaked electric field shown in Fig.~\ref{fig:evl_sketch}.  The relative signal minimum near $y=700\ \mu$m (see Fig.~\ref{fig:dj44}) corresponds to the minimum of the electric field $z$-component, $E_z$, where both electrons and holes travel only short distances before trapping.  This small separation induces only a small signal on the n$^+$ side of the detector.  At larger values of $y$, $E_z$ increases causing the electrons drift back into the minimum where they are likely to be trapped.  However, the holes drift into the higher field region near the p$^+$ implant and are more likely to be collected.  The net induced signal on the n$^+$ side of the detector therefore increases and creates the local maximum seen near $y=900\ \mu$m.

\begin{figure}[thb]
  \begin{center}
    \resizebox{0.75\linewidth}{!}{\includegraphics{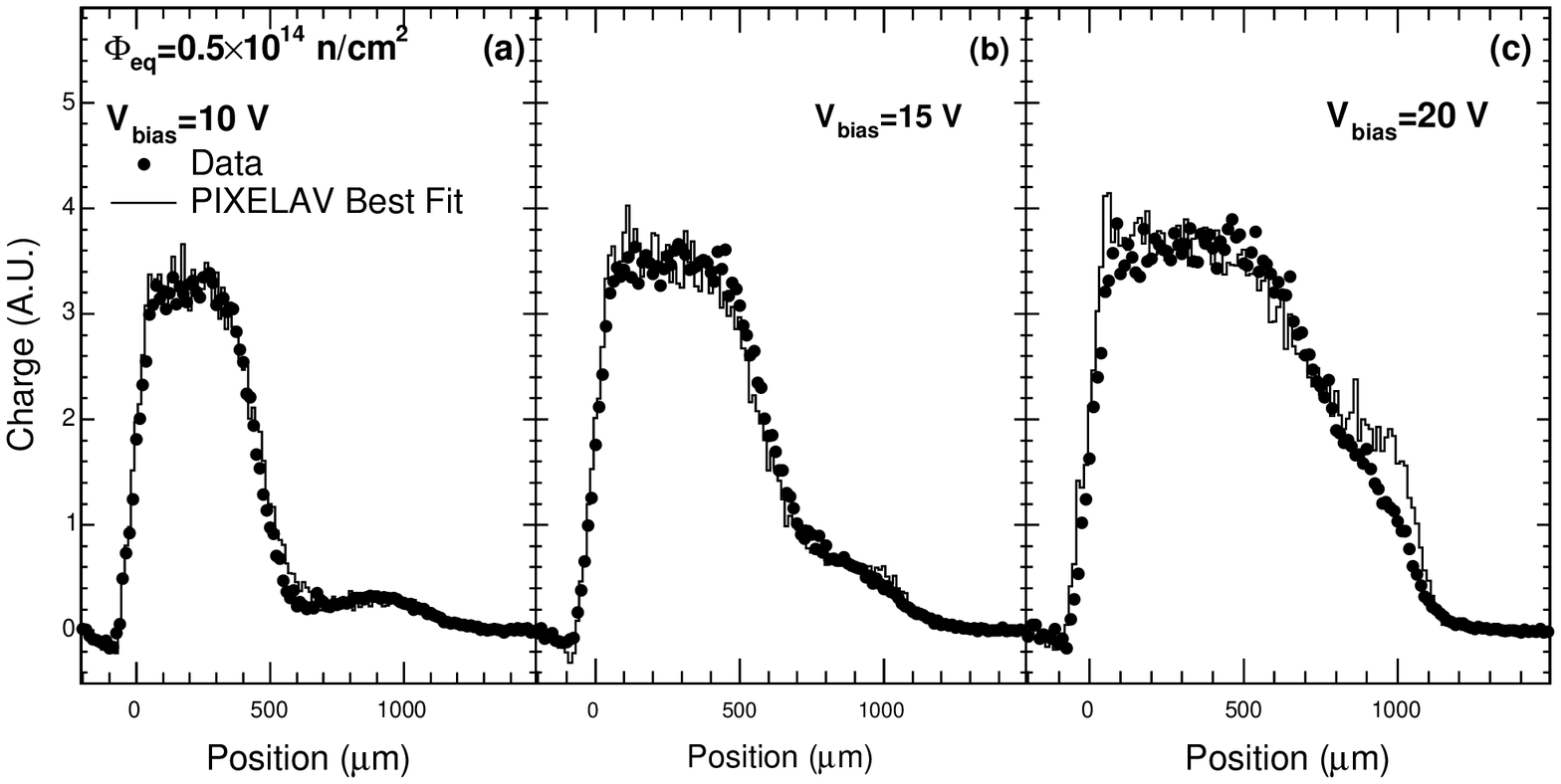}}
    \resizebox{\linewidth}{!}{\includegraphics{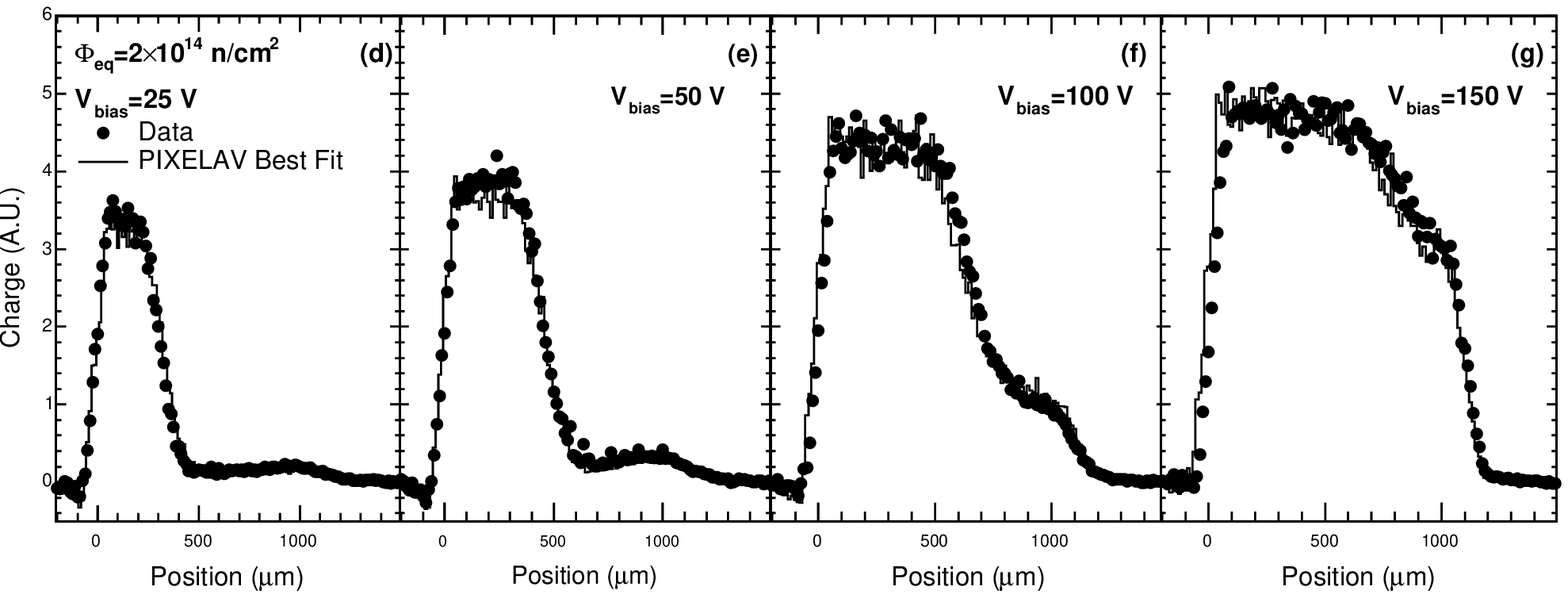}}
  \caption{Measured (full dots) and simulated (histogram) charge collection profiles for sensors irradiated 
to fluences of $\Phi = 0.5\times10^{14}$~n$_{\rm eq}$/cm$^2$ (a-c) and of $\Phi = 2\times10^{14}$~n$_{\rm eq}$/cm$^2$ (d-g),
at $T=-10^\circ$C and several bias voltages.}
  \label{fig:summary_2N} 
  \end{center}
\end{figure}

The charge collection profiles at $T=-10^\circ$C for sensors irradiated to fluences of
$\Phi = 0.5\times10^{14}$~n$_{\rm eq}$/cm$^2$ and $\Phi = 2\times10^{14}$~n$_{\rm eq}$/cm$^2$ 
and operated at several bias voltages are presented in Fig.~\ref{fig:summary_2N}(a-c)
and Fig.~\ref{fig:summary_2N}(d-g), respectively. The measured profiles,
shown as solid dots, are compared to the simulated profiles, shown as histograms.  Note that the ``wiggle'' is present at low bias even at $\Phi = 0.5\times10^{14}$~n$_{\rm eq}$/cm$^2$ which is just above the ``type-inversion'' fluence.  This suggests that a doubly peaked field is present even at rather small fluences.

%
%

The double junction model can provide a reasonable description of the lower fluence charge collection profiles using the parameters obtained with the fitting procedure shown in 
Table~\ref{tab:model_values}.  We observe that the donor trap concentration increases more rapidly with fluence than does the acceptor trap concentration. The ratio between acceptor
and donor trap concentrations is 0.76 at the lowest fluence and decreases 
to 0.40 at $5.9\times$10$^{14}$~n$_{\rm eq}$/cm$^2$. In addition, the fits exclude 
a linear dependence of the trap concentrations with the irradiation fluence.
At $\Phi = 5.9\times$10$^{14}$~n$_{\rm eq}$/cm$^2$ the cross section ratio 
$\sigma_h/\sigma_e$ is set to 0.25 for both donor and acceptor traps
while at lower fluences we find $\sigma^{A}_h/\sigma^{A}_e = 0.25$ and 
$\sigma^{D}_h/\sigma^{D}_e = 1$ for the acceptor and donor traps, respectively.  The simulated leakage current is approximately linear in fluence, but the ratio $N_A/N_D$ is clearly not constant.  This may be a consequence of the quadratic fluence scaling of one or more di-vacancy states or it may reflect the fact that the two trap model with the particular choice of activation energies does not accurately model the dependence of the trap occupancies on leakage current.  The allowed $N_D$-$\sigma_e$ parameter spaces for the lower fluence models are much more constrained than in the $\Phi$=$5.9\times$10$^{14}$~n$_{\rm eq}$/cm$^2$ case and predict the expected leakage current.
%
%

\begin{table}[h]
\begin{center}
\begin{tabular}{lccc}
\hline
$\Phi$  [$10^{14}$~n$_{\rm eq}$cm$^{-2}$]  & 0.50  & 2.0  & 5.9 \\
\hline
$N_A$ [$10^{14}$~cm$^{-3}$] & 1.9 & 6.8 & 16  \\
$N_D$ [$10^{14}$~cm$^{-3}$]  & 2.5 & 10 & 40 \\
$\sigma^{A/D}_e$ [$10^{-15}$~cm$^2$] & 6.60 & 6.60 & 6.60 \\
$\sigma^A_h$ [$10^{-15}$~cm$^2$] & 1.65 & 1.65 & 1.65 \\
$\sigma^D_h$ [$10^{-15}$~cm$^2$] & 6.60 & 6.60 & 1.65 \\\hline
\end{tabular}
\end{center}
\caption{Double trap model parameters extracted from the fit to the data.}
\label{tab:model_values}
\end{table}

\begin{figure}[t]
  \begin{center}
    \resizebox{\linewidth}{!}{    
      \includegraphics[angle=90]{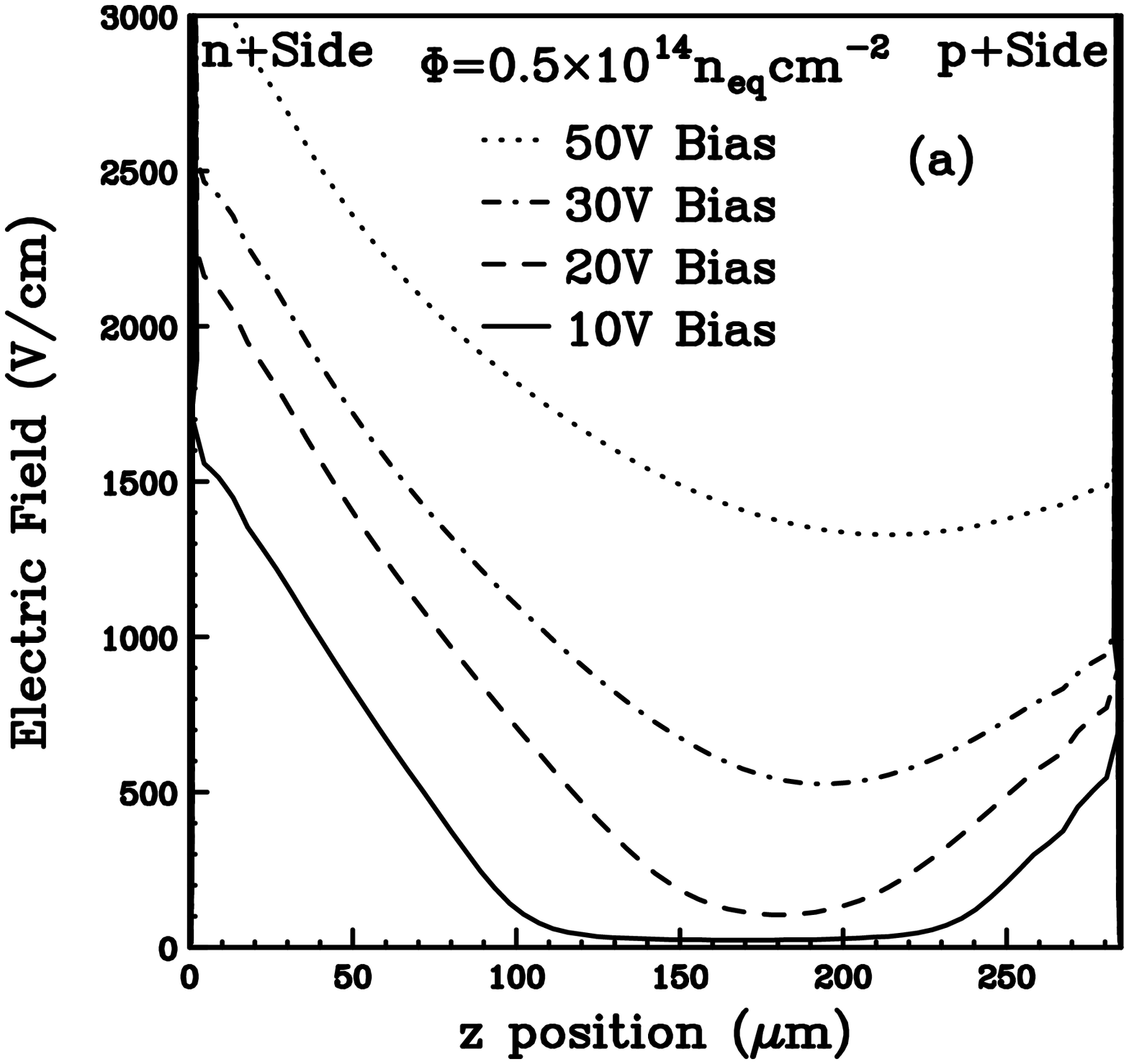} 
      \includegraphics[angle=90]{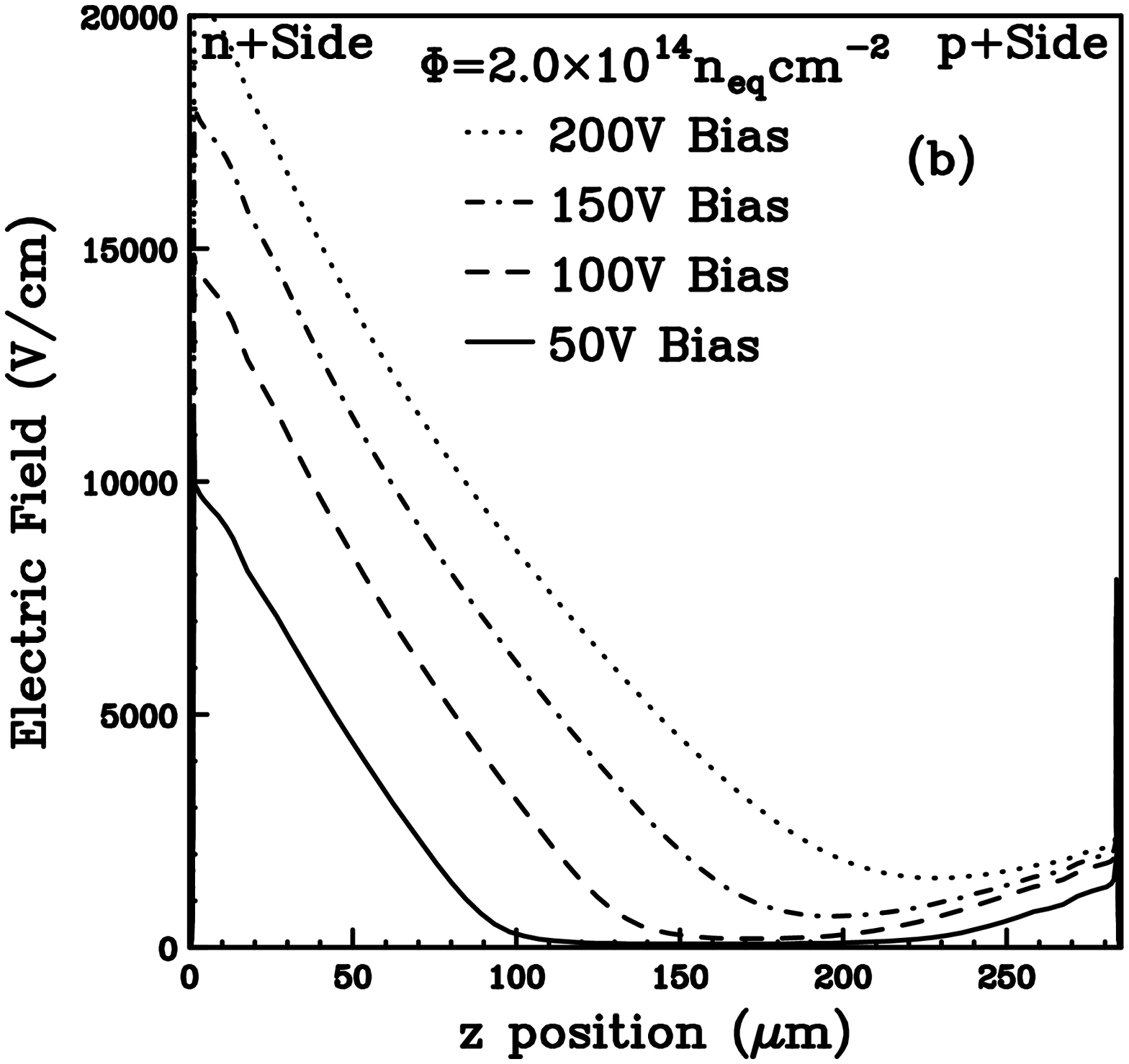} 
	      }
  \end{center}
  \caption{The $z$-component of the simulated electric field at $T=-10^\circ$C resulting from the model best fit is shown as a function of $z$ for a sensor irradiated to fluences of $\Phi = 0.5\times10^{14}$~n$_{\rm eq}$/cm$^2$ (a) and $\Phi = 2\times10^{14}$~n$_{\rm eq}$/cm$^2$ (b).}
  \label{fig:E_z_profile} 
\end{figure}

\begin{figure}[b]
  \begin{center}
    \resizebox{0.75\linewidth}{!}{\includegraphics[angle=90]{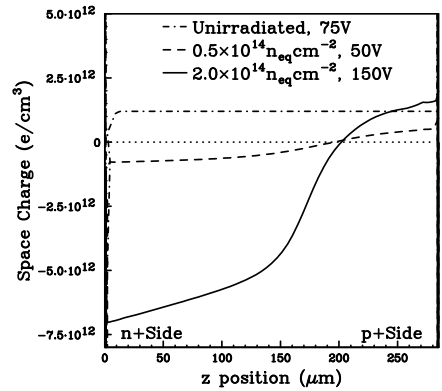}}
  \caption{The simulated space charge density at $T=-10^\circ$C as a function of the $z$ coordinate for fluences of $\Phi = 0.5\times10^{14}$~n$_{\rm eq}$/cm$^2$ and $\Phi = 2\times10^{14}$~n$_{\rm eq}$/cm$^2$.  }
  \label{fig:spacecharge} 
  \end{center}
\end{figure}

The $z$-component of the simulated electric field, $E_z$, is plotted as a function of $z$ in Fig.~\ref{fig:evl_sketch} for $\Phi = 5.9\times10^{14}$~n$_{\rm eq}$/cm$^2$ and in Fig.~\ref{fig:E_z_profile} for $\Phi = 0.5\times10^{14}$~n$_{\rm eq}$/cm$^2$ and $\Phi = 2\times10^{14}$~n$_{\rm eq}$/cm$^2$.  At $\Phi = 5.9\times10^{14}$~n$_{\rm eq}$/cm$^2$, the field profile has a minimum near the midplane of the detector which shifts toward the p+ implant at lower fluences.  The field has maxima at the detector implants as discussed in Section~\ref{sec:simulation}.  Figure~\ref{fig:E_z_profile}(a) shows that a 
doubly peaked electric field is necessary to describe the measured charge collection profiles 
even at the lowest measured fluence which is just beyond the ``type inversion point''.
The dependence of the space charge density upon the $z$ coordinate is shown in Figures~\ref{fig:evl_sketch} and \ref{fig:spacecharge}.
Before irradiation the sensor is characterized by a constant and positive space charge density of $1.2\times10^{12}$~cm$^{-3}$
across the sensor bulk.
After a fluence of $0.5\times10^{14}$~n$_{\rm eq}$/cm$^2$ the device shows a negative
space charge density of about $-1\times10^{12}$~cm$^{-3}$ for about 70\% of its thickness, a compensated
region corresponding to the $E_z$ minimum and a positive space charge density close to the backplane.  The space charge density and electric field near the p+ implant increase with fluence.
The space charge density is not linear in $z$ due to the variation of the carrier drift velocities with the electric fields.
\section{Temperature dependence\label{sec:temperature}}

\begin{figure}[thb]
  \begin{center}
    \resizebox{0.67\linewidth}{!}{\includegraphics{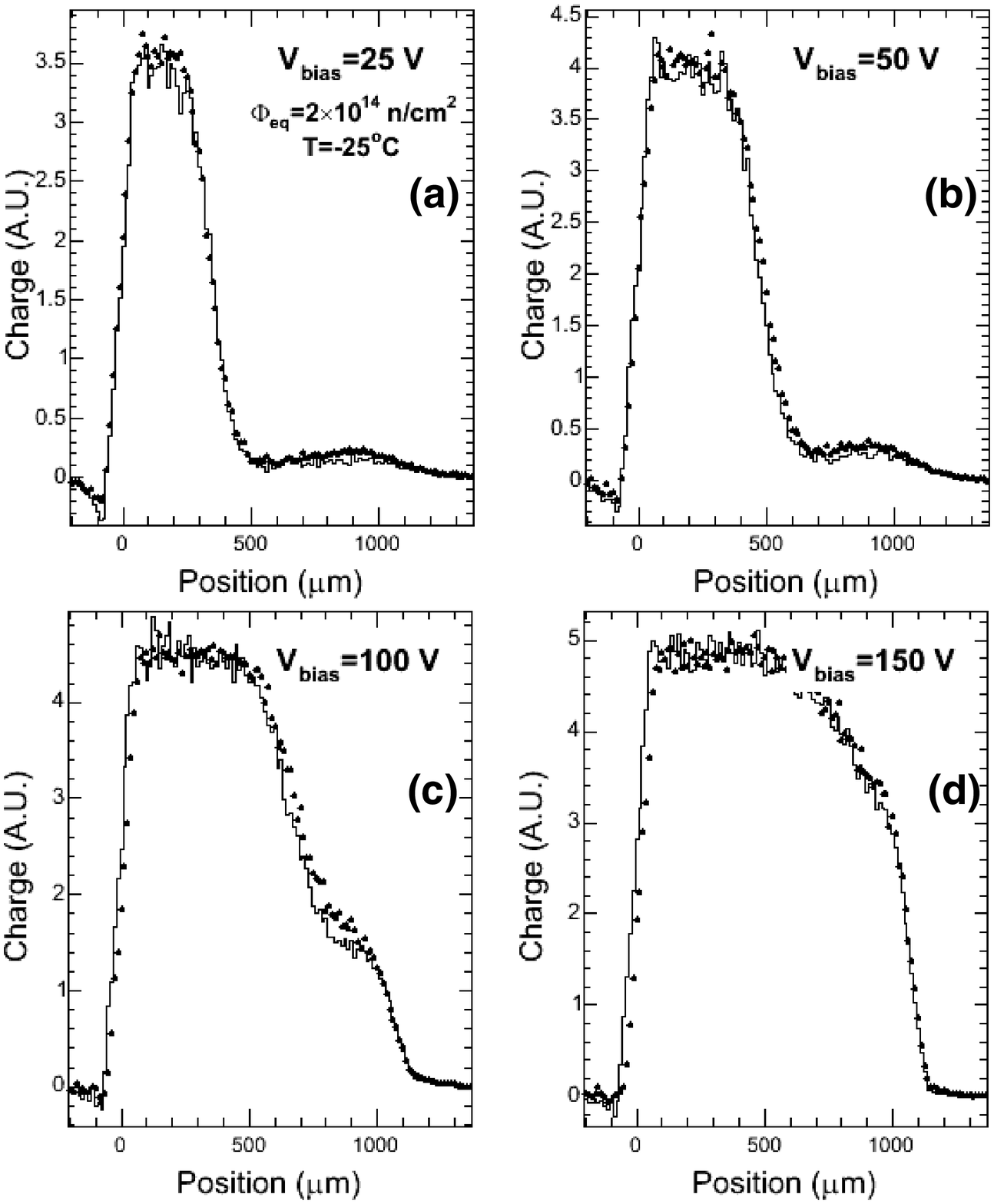}}
    \resizebox{\linewidth}{!}{\includegraphics{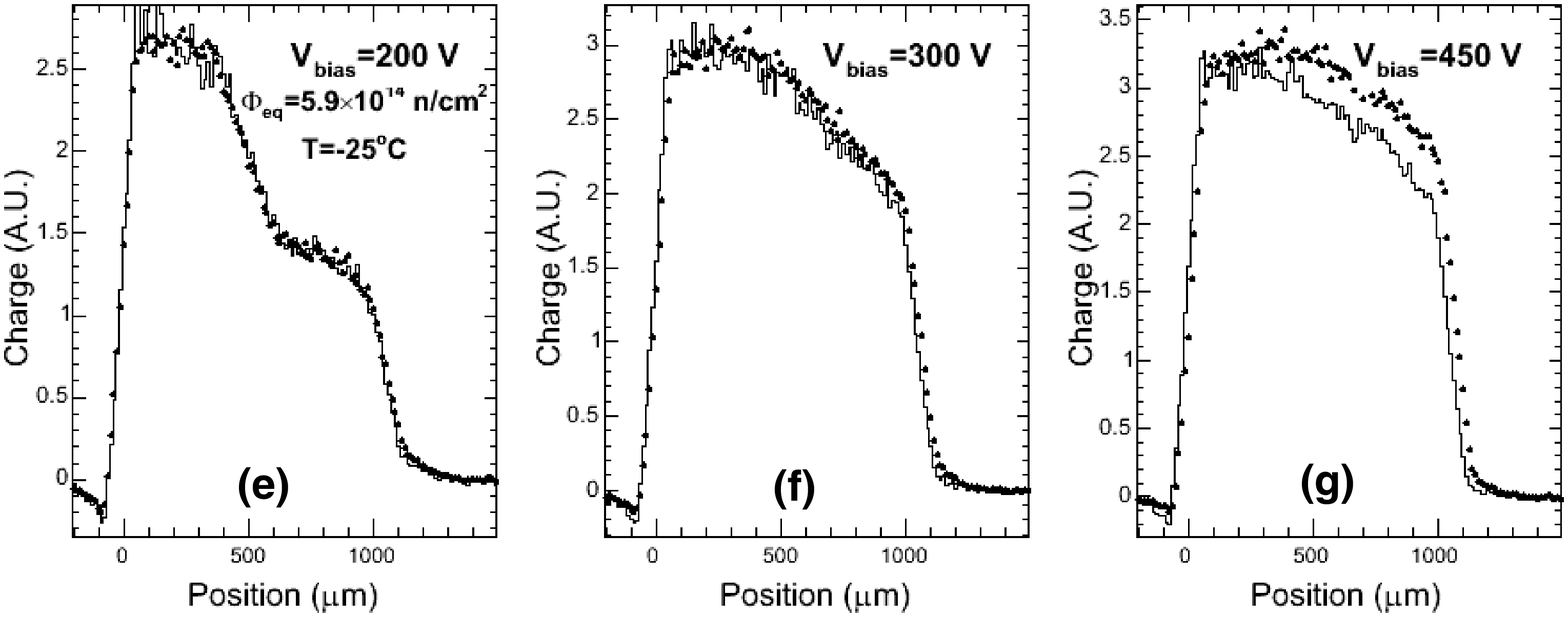}}
  \caption{Measured (full dots) and simulated (histogram) charge collection profiles at $T=-25^\circ$C and several bias voltages for sensors irradiated to fluences of $\Phi = 2.0\times10^{14}$~n$_{\rm eq}$/cm$^2$ (a-d) and of $\Phi = 5.9\times10^{14}$~n$_{\rm eq}$/cm$^2$ (e-g).}
  \label{fig:summary_T} 
  \end{center}
\end{figure}

The temperature dependence of the charge collection profiles was studied by accumulating data at $T=-25^\circ$C.  The {\sc pixelav} simulation includes temperature dependent mobilities, diffusion, and trapping rates.  The {\sc tcad} calculation of the electric field map is also based upon temperature dependent quantities including the bandgap energy and SRH lifetimes.  The $T=-25^\circ$C charge collection profiles for the $\Phi = 2.0\times10^{14}$~n$_{\rm eq}$/cm$^2$ and $\Phi = 5.9\times10^{14}$~n$_{\rm eq}$/cm$^2$ sensors are compared with the simulation in Fig.~\ref{fig:summary_T}.  It is clear that the simulation correctly tracks the temperature-dependent variations in the measured profiles.  

\begin{figure}[hbt]
  \begin{center}
    \resizebox{0.75\linewidth}{!}{\includegraphics{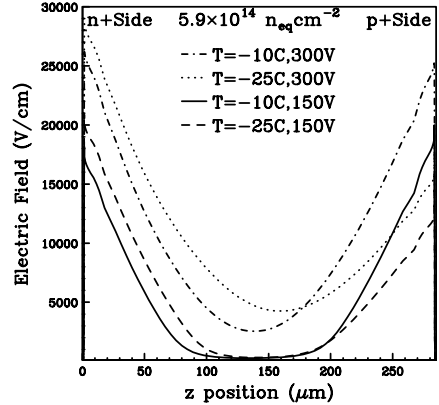}}
  \caption{The simulated $z$-component of the electric field as a function of the $z$ coordinate at the $\Phi = 5.9\times10^{14}$~n$_{\rm eq}$/cm$^2$ fluence for temperatures  $T=-10^\circ$C and $T=-25^\circ$C.  The field profiles are shown for bias voltages of 150V and 300V.}
  \label{fig:EvsT} 
  \end{center}
\end{figure}

The effect of temperature on the $z$-component of the simulated electric field at  $\Phi = 5.9\times10^{14}$~n$_{\rm eq}$/cm$^2$ is shown in Fig.~\ref{fig:EvsT} for bias voltages of 150V and 300V.  It is clear that decreasing the temperature also decreases the fields on the p+ side of the sensor and increases them on the n+ side.

\section{Conclusions\label{sec:conclusions}}
%
In this paper we have shown that doubly peaked electric fields are necessary to describe grazing-angle charge collection measurements of irradiated silicon pixel sensors.  A model of irradiated silicon based upon two defect levels with opposite charge states and the trapping of charge carriers can
be tuned to produce a good description of the measured charge collection profiles
in the fluence range from $0.5\times$10$^{14}$~n$_{\rm eq}$/cm$^2$
to $5.9\times$10$^{14}$~n$_{\rm eq}$/cm$^2$.  The model correctly predicts the variation in the profiles as the temperature is changed from $-10^\circ$C to $-25^\circ$C.

The doubly peaked electric
field profiles have maxima near the implants and
minima near the detector midplane. This corresponds to  
negative space charge density near the n$^+$ implant and
and positive space charge density near the p$^+$ backplane. 
We find that it is necessary to decrease the ratio of acceptor concentration
to donor concentration as the fluence increases. This causes the electric
field profile to become more symmetric as the fluence increases.  The effect of decreasing the temperature has the opposite effect of suppressing the fields on the p+ side of the sensor and increasing them on the n+ side.

The measured charge collection profiles of irradiated sensors are inconsistent with the linearly-varying electric fields predicted by the usual description based upon a uniform effective doping density.  This suggests that the correctness and the physical significance 
of effective doping densities determined from capacitance-voltage measurements
are quite unclear. In addition, we remark
that the notion of partly depleted silicon sensors after irradiation 
is inconsistent with the measured charge collection profiles and 
with the observed doubly peaked electric fields.

The charge-sharing behavior and resolution functions of many detectors are sensitive to the details of the internal electric field.  A known response function is a key element of any reconstruction procedure.  A working effective model will permit the detailed response of these detectors to be tracked as they are irradiated in the next generation of accelerators.

\section*{Acknowledgments}
We gratefully acknowledge Silvan Streuli from ETH Zurich and Fredy Glaus from PSI for
their immense effort with the bump bonding, Federico Ravotti, Maurice Glaser and Michael Moll from CERN for
carrying out the irradiation, Kurt B\"osiger from the Z\"urich University workshop for the mechanical
construction, Gy\"orgy Bencze and Pascal Petiot from CERN for the H2 beam line support
and, finally, the whole CERN-SPS team.  This work was supported in part by NSF grant PHY-0457374.


\bibliographystyle{elsart-num}    



\end{document}